\documentstyle[prl,twocolumn,aps]{revtex}
\input psfig
\begin{document}
\draft

%----------------------User's Commands----------------------------
 \newcommand{\mytitle}[1]{
 \twocolumn[\hsize\textwidth\columnwidth\hsize
 \csname@twocolumnfalse\endcsname #1 \vspace{1mm}]}
%------------------------------------------------------------------------

\mytitle{
\title{Nonperturbative analysis of coupled quantum dots in a phonon bath}

\author{Markus Keil$^1$ and Herbert Schoeller$^1$}

\address{
$^1$ Institut f\"ur Theoretische Physik, Rheinisch-Westf\"alische
Technische Hochschule Aachen, 52056 Aachen, Germany
}

\date{\today}

\maketitle

\begin{abstract}
Transport through coupled quantum dots in a phonon bath is studied using
the recently developed real-time renormalization-group
method. Thereby, the problem can be treated beyond perturbation theory
regarding the complete interaction. A
reliable solution for the stationary tunnel current is obtained for
the case of moderately strong couplings of
the dots to the leads and to the phonon bath. Any other parameter is
arbitrary, and the
complete electron-phonon interaction is taken into
account. Experimental results are
quantitatively reproduced by taking into account a finite extension of
the wavefunctions within the dots. Its dependence on the energy
difference between the dots is derived.
\end{abstract}
}

{\it Introduction}.
Today quantum dot systems allow a detailed study of many physical
phenomena, like Coulomb blockade~\cite{raikh,tarucha}, Kondo
effects~\cite{cronenwett,wiel,sasaki}, interference
effects~\cite{wiel}. As in these structures quantum states can be
manipulated, they also may have an application in future quantum
gates~\cite{loss}. 
Quantum dot systems can typically be characterized by only a few parameters,
which are experimentally controllable. Theoretically, these systems
may be described by basic models, which capture the essential
physics and can be investigated using standard methods of many-particle
theory. However, if one deals with out-of-equilibrium situations and
strong coupling, a
theoretical analysis becomes difficult.

Such an out-of-equilibrium problem was recently studied, when the
stationary tunnel current through a double quantum dot in a
phonon bath was measured~\cite{fujisawa}. There the influence of the 
phonon environment was examined at low temperature. The thermal energy
of the environment
is always a source of unwanted transitions in quantum dot
devices. Even at zero temperature spontaneous emission of phonons
gives rise to inelastic transitions, i.e. they occur between dot
states of nonequal energy. In the experiment in Ref.~\cite{fujisawa} the
inelastic contribution to the tunnel current through the double dot
was studied. A first
theoretical interpretation of the experimental results focused on the
interaction
of the dots with the phonons, which is analogous to the spin-boson
model~\cite{brandes}. They accounted for the coupling to the leads
perturbatively and studied the electron-phonon problem using an approximation,
which corresponds to the
noninteracting blip approximation (NIBA) for the spin-boson
model~\cite{leggett-etal}. Thereby a better understanding of the shape of the
inelastic current
spectrum was achieved. It was shown that the interference
of phonons interacting with the electron densities at the two dots
leads to an oscillating structure in the current spectrum.
However, a quantitative comparison
with the experiment has not yet been possible. Especially, the
unexpectedly large
inelastic current of the experiment could not be explained.

In this paper we present a reliable solution for the
stationary state of the double dot system. Since we
deal with a nonequilibrium situation, where both the coupling to the
phonon reservoir and the coupling to the leads have to be considered
as strong, we use the
recently developed real-time
renormalization-group (RTRG) method. It has successfully been applied
to both equilibrium~\cite{koenig-hs,mk-hs_pol} and nonequilibrium
problems~\cite{hs-koenig,mk-hs_sbm}, including the spin-boson
model~\cite{mk-hs_sbm,mk-hs_sbmproc}.
Applying this approach to the double dot system both the electron and
the phonon reservoirs are integrated out in a RG procedure,
i.e. beyond perturbation theory. Thereby, the external voltage is
accounted for properly, and the level broadening induced by
the coupling to the leads is included in this method. Thus, we
can also consider the case,
where the energy difference between the dots is of the same order as
the external voltage. Furthermore, we do not have to include an
additional cutoff parameter, which simulates the level broadening
generated by the leads. Moreover,
this formalism may treat any form of dot-phonon interaction. Therefore, we
are able to account for the full electron-phonon interaction, i.e.
including interaction terms, which involve a tunneling between the dots
(``offdiagonal interaction terms'').
In the quantitative analysis
the offdiagonal interaction terms lead to a strong
dependence of the current on the extension of the
wavefunctions within one dot. We find that the variation of this extension
with the energy difference $\epsilon$ between the dot levels has to be
accounted for. By fitting the result for the current with the
experimental data, we obtain the width of the electron density within one
dot as a function of $\epsilon$.
Furthermore, for the physically relevant situation, where the
coupling of the dots to the leads and to the phonons is only
moderately large, the RTRG approach does not deal with any
parameter restriction. This is in contrast to the NIBA, which is valid
only for sufficiently high temperature.

{\it Model}.
Our model consists of two coupled quantum dots ($l$ and $r$,
respectively). Each dot is coupled to an electron reservoir with the
chemical potentials $\mu_l$ and $\mu_r$, see Fig.~\ref{ddfig}. We consider
the case realized
in the experiment~\cite{fujisawa}, where
the external voltage $V=\mu_l-\mu_r$ is much smaller than the Coulomb
charging energy $U$. Thus, due to Coulomb blockade the double dot
cannot be charged with more than one additional electron. In the
experiment~\cite{fujisawa} a strong magnetic field was applied
perpendicular to the dots. Thus, we assume spin polarization here and
omit the spin index. 
We denote the many-particle ground states, where an additional electron is in
the left (right) dot, by $|l\rangle$ ($|r\rangle$) and neglect any
excited states.
Therefore, together with the uncharged ground state $|0\rangle$,
there are only three possible states of the double dot. 
The total Hamiltonian $\bar H$ for the system can be written as a sum
of the dot
Hamiltonian, the contributions of the electron reservoirs and the
phonon bath, and the interaction parts stemming from the coupling to
the leads and the electron-phonon interaction:
\begin{equation}
\label{hsumeq}
\bar H = H_d + H_{res} + H_{ph} + H_{e-res} + H_{e-ph}\,.
\end{equation}
The dot Hamiltonian $H_d$ reads
\begin{equation}
\label{hdeq}
H_d = \epsilon_l |l\rangle\langle l| + \epsilon_r |r\rangle\langle r|
+ T_c \left(|l\rangle\langle r| + |r\rangle\langle l|\right)\,,
\end{equation}
where $\epsilon_l$ ($\epsilon_r$) are the ground state energies of
$|l\rangle$ ($|r\rangle$) and the coupling between the dots is
described by the tunnel matrix element $T_c$.
The reservoir contributions are given by
\begin{eqnarray}
\label{hreq}
H_{res} &=& \sum_k \epsilon_k c^{\dagger}_kc_k + \sum_k \epsilon_k
d^{\dagger}_kd_k\,,\\
\label{hpeq}
H_{ph} &=& \sum_q \omega_q a^{\dagger}_qa_q\,.
\end{eqnarray}
Here, the operators $c^{\dagger}_k$ ($c_k$) create (annihilate) an
electron with the energy $\epsilon_k$ in the left lead, whereas the
creation (annihilation) operators $d^{\dagger}_k$ ($d_k$) refer to the
right electron reservoir. Analogously, $a^{\dagger}_q$ ($a_q$) create
(annihilate) a phonon with the wavevector $\vec q$ and the frequency
$\omega_q$. Here and
throughout the paper we set $\hbar=1$. 
The double dot is coupled to the external leads by the parameters $V_k$
and $W_k$:
\begin{eqnarray}
H_{e-res} &=& \sum_k \left(V_k c_k|l\rangle\langle 0| + V^*_k
|0\rangle\langle l|c^{\dagger}_k\right)\nonumber\\
&& + \sum_k \left(W_k d_k|r\rangle\langle
0| + W^*_k |0\rangle\langle r|d^{\dagger}_k\right)\,.
\label{hereq}
\end{eqnarray}
The electron-phonon interaction consists of a diagonal part, which is
characterized by the coupling constants $\alpha_q$ and $\beta_q$, and
an offdiagonal contribution with the parameter $\gamma_q$:
\begin{eqnarray}
H_{e-ph} &=& \sum_q (\alpha_q|l\rangle\langle l| +
\beta_q|r\rangle\langle r|)(a^{\dagger}_q + a_{-q})\nonumber\\
&& + \sum_q \gamma_q(|l\rangle\langle r| +
|r\rangle\langle l|)(a^{\dagger}_q + a_{-q})\,.
\label{hepeq}
\end{eqnarray}
The above interaction coefficients are given by~\cite{brandes}
\begin{eqnarray}
\label{alphaqeq}
\alpha_q &=& \lambda_q \langle l|e^{i \vec q \vec x}|l\rangle\,,\\
\label{betaqeq}
\beta_q &=& \lambda_q \langle r|e^{i \vec q \vec x}|r\rangle\,,\\
\label{gammaqeq}
\gamma_q &=& \lambda_q \langle l|e^{i \vec q \vec x}|r\rangle\,,
\end{eqnarray}
where $\lambda_q$ is the matrix element for the interaction of 2DEG
electrons and phonons. The phonons are assumed to be three-dimensional
acoustical phonons~\cite{fujisawa}. It then follows for the
interaction~\cite{brandes}
\begin{equation}
\label{lambdaqeq}
|\lambda_q|^2 = g \frac{\pi^2 c_s^2}{V|\vec q|}\,,
\end{equation}
and the dispersion reads
\begin{equation}
\label{omegaqeq}
\omega_q = c_s |\vec q|\,.
\end{equation}
Here, we introduced $c_s$ as the speed of sound in the medium, $V$
as the volume of the crystal, and the dimensionless coupling constant
$g$~\cite{brandes}.
For the evaluation of Eqs.~(\ref{alphaqeq}) - (\ref{gammaqeq}) we
model the electron densities $\rho_l(\vec x)$ ($\rho_r(\vec x)$) within one
dot by Gaussians, which are peaked around the dot positions $x_l$
($\vec x_r = \vec
x_l + \vec d$) with a width $|\Delta\vec x|=\sqrt{3/2}\,\sigma$:
\begin{equation}
\label{rholreq}
\rho_{l(r)}(\vec x) = \left(\frac{1}{\pi\sigma^2}\right)^{3/2}
e^{-\frac{(\vec x - \vec x_{l(r)})^2}{\sigma^2}}\,.
\end{equation}
The finite width $\sigma$ leads to a high-energy cutoff $D=c_s/\sigma$ for the
coefficients $\alpha_q$, $\beta_q$ and $\gamma_q$. We include this
cutoff in an exponential form, so that we end up with the following
interaction coefficients:
\begin{eqnarray}
\label{alphaq1eq}
\alpha_q &=& \lambda_q e^{i \vec q \vec x_l} e^{-\frac{c_s|\vec q|}{2D}}\,,\\
\label{betaq1eq}
\beta_q &=& \lambda_q e^{i \vec q \vec x_r} e^{-\frac{c_s|\vec q|}{2D}}\,,\\
\label{gammaq1eq}
\gamma_q &=& \lambda_q e^{i \vec q (\frac{\vec x_l + \vec x_r}{2})}
e^{-\frac{|\vec d|D}{2c_s}} e^{-\frac{c_s|\vec q|}{2D}}\,.
\end{eqnarray}
A simple form of the Hamiltonian, which shows the analogy with the
spin-boson model, is obtained by shifting the bosonic field
operators. One introduces the unitary transformation
\begin{equation}
\label{ueq}
U =
\exp\left[\sum_q\left(\frac{\alpha_q+\beta_q}{2\omega_q}a^{\dagger}_q
- \frac{\alpha^*_q+\beta^*_q}{2\omega_q}a_q\right)\right]\,,
\end{equation}
so that
\begin{equation}
\label{bqeq}
Ua_qU^{\dagger} = a_q - \frac{\alpha_q + \beta_q}{2\omega_q}\,.
\end{equation}
Thus, our final Hamiltonian $H=U\bar HU^{\dagger}$ reads
\begin{eqnarray}
\label{hsumfinaleq}
H &=& H_0 + H_B + H_V\,,\\
H_0 &=& \frac{\epsilon}{2} \left(|l\rangle\langle l| -
|r\rangle\langle r|\right)\nonumber\\
&& + T_c^{\rm eff}
\left(|l\rangle\langle r| + |r\rangle\langle l|\right)\nonumber\\
&& + E
\left(|l\rangle\langle l| + |r\rangle\langle r| - |0\rangle\langle 0|\right)
\label{h0finaleq}
\,,\\
\label{hbfinaleq}
H_B &=& \sum_k \epsilon_k c^{\dagger}_kc_k + \sum_k \epsilon_k
d^{\dagger}_kd_k + \sum_q \omega_q a^{\dagger}_qa_q\,,\\
\label{hvfinaleq}
H_V &=& \sum_{\mu}:g_{\mu}j_{\mu}:\,,
\end{eqnarray}
where we have used Eqs.~(\ref{omegaqeq}) and (\ref{alphaq1eq}) -
(\ref{gammaq1eq}). Furthermore, for simplicity we have
set $(\epsilon_l+\epsilon_r)/2=0$ and have introduced the parameters
\begin{eqnarray}
\label{epseq}
\epsilon &=& \epsilon_l-\epsilon_r\,,\\
\label{Tceq}
T_c^{\rm eff} &=& T_c -
2g\omega_d e^{-\frac{D}{2\omega_d}}\arctan\frac{D}{2\omega_d}\,,\\
\label{Eeq}
E &=& -\frac{g}{4} \left(D +
\omega_d\arctan\frac{D}{\omega_d}\right)
\end{eqnarray}
and
\begin{equation}
\label{omdeq}
\omega_d = \frac{c_s}{|\vec d|}\,.
\end{equation}
Thus, the tunnel
amplitude $T_c$ has to be replaced by a smaller effective $T_c^{\rm eff}$,
which is due to the offdiagonal electron-phonon interaction. One
already recognizes that the reduction of $T_c$ strongly depends on the
width of the electron
densities $\sigma=|\vec d|\omega_d/D$.
Finally, in view of the RTRG method, we have written the interaction part $H_V$
as normal ordered products of local (dot) operators $g_{\mu}$ and environmental
operators $j_{\mu}$.
They are defined by
\begin{eqnarray}
\label{gb1eq}
g_{b_1} &=& \frac{1}{2}\left(|l\rangle\langle l| - |r\rangle\langle
r|\right)\,,\\
\label{jb1eq}
j_{b_1} &=& \sum_q \left(\alpha_q - \beta_q\right) \left(a^{\dagger}_q +
a_{-q}\right)\,,\\
\label{gb2eq}
g_{b_2} &=& \left(|l\rangle\langle r| + |r\rangle\langle
l|\right)\,,\\
\label{jb2eq}
j_{b_2} &=& \sum_q \gamma_q \left(a^{\dagger}_q +
a_{-q}\right)\,,\\
\label{gb3eq}
g_{b_3} &=& -\frac{1}{2} |0\rangle\langle 0|\,,\\
\label{jb3eq}
j_{b_3} &=& \sum_q \left(\alpha_q + \beta_q\right) \left(a^{\dagger}_q +
a_{-q}\right)\,,\\
\label{gjleq}
g_{+l} &=& g_{-l}^{\dagger} = |l\rangle\langle 0| \quad,\qquad j_{+l}
= j_{-l}^{\dagger} = \sum_k V_k c_k\,,\\
\label{gjreq}
g_{+r} &=& g_{-r}^{\dagger} = |r\rangle\langle 0| \quad,\qquad j_{+r}
= j_{-r}^{\dagger} = \sum_k W_k d_k\,.\\
\end{eqnarray}
Therefore, the interaction index $\mu$ runs over the bosonic indices
$b_1,b_2,b_3$ and the fermionic ones $+l,-l,+r,-r$.
From Eqs.~(\ref{hsumfinaleq}) - (\ref{hvfinaleq}) the spin-boson model is
recovered by omitting
the electron reservoirs and the interaction with them, excluding
the state $|0\rangle$ and accounting only for the bosonic interaction
$b_1$. The latter corresponds to neglecting the offdiagonal
electron-phonon ($b_2$) interaction.

{\it RTRG approach}.
The RTRG approach is based on the formally exact kinetic equations
\begin{eqnarray}
\label{iteq}
\langle I\rangle (t) &=& {\rm Tr}_0
\left[\int_0^tdt'\, \Sigma_I(t-t')p(t')\right]\,,\\
\dot{p}(t) + iL_0p(t) &=& \int_0^tdt'\, \Sigma(t-t')p(t')\,,
\label{pteq}
\end{eqnarray}
where $\langle I\rangle(t)$ denotes the time-dependent expectation
value of the current through the double dot, and $p(t)$ is the
time-dependent reduced density matrix, which is the trace over the bath
degrees of freedom ${\rm Tr}_B$ of the full density matrix. In contrast, ${\rm
Tr}_0$ denotes the trace over the local (dot) degrees of freedom. The
bath degrees of freedom enter the equations via the
integral kernels $\Sigma$ and $\Sigma_I$, which are defined by sums
over irreducible diagrams, see Ref.~\cite{hs} for details.
Introducing the Laplace transforms $f(z)=\int_0^{\infty}dt\,
e^{izt}f(t)$
of the time-dependent functions $f(t)$ and using the identity
$\lim_{t\rightarrow\infty}f(t)=-i\lim_{z\rightarrow 0}zf(z)$ leads to
the following expression for the stationary tunnel
current $I_{st}$:
\begin{equation}
\label{isteq}
I_{st} = {\rm Tr}_0\left[\Sigma_I(z=0)p_{st}\right]\,,
\end{equation}
where the stationary reduced density matrix $p_{st}$ is determined by
\begin{equation}
\label{psteq}
\left(L_0 + i\Sigma(z=0)\right)p_{st} = 0\,.
\end{equation}
In the following we outline only the main steps of the (i) derivation
of Eqs.~(\ref{iteq}) and (\ref{pteq}), and of the (ii) RTRG technique, by
which $\Sigma$ and
$\Sigma_I$ are calculated.

(i) The quantity of interest is the current through the double dot. It is
given by the expectation value of the operator
\begin{equation}
\label{idefeq}
I = ie \sum_k \left(V^*_k|0\rangle\langle l|c^{\dagger}_k -
V_kc_k|l\rangle\langle 0|\right)
\end{equation}
with $e$ being the elementary charge.
For the calculation of $\langle I\rangle(t)$ we introduce the
Liouvillian $L=[H,\cdot]_-$ and the
superoperator $A_I=[\frac{i}{2}I,\cdot]_+$, where
$[\cdot,\cdot]_{-(+)}$ denotes the (anti)commutator. The expectation
value of the current can then be written as
\begin{equation}
\label{it1eq}
\langle I\rangle (t) = -i {\rm Tr} \left[A_I e^{-iLt} p(0)\rho_B^{eq}\right]\,.
\end{equation}
Here, we assumed a factorized density matrix at $t=0$, where $p(0)$ is
the initial dot density matrix and $\rho_B^{eq}$ is the equilibrium
distribution of the electron reservoirs and the phonon bath.
To evaluate the above expression the propagator $\exp{(-iLt)}$ is
expanded in the interaction 
part $L_V=[H_V,\cdot]$. According to Eq.~(\ref{hvfinaleq}) this
can be written as
\begin{equation}
\label{lvlepeq}
L_V = \sum_{p\mu}:G^p_{\mu}J^p_{\mu}:\,,
\end{equation}
where the index $p=\pm$ denotes, whether the interaction takes place on
the forward or backward propagator, i.e.
\begin{eqnarray}
\label{gpeq}
G^+_{\mu} &=& g_{\mu}\,\cdot \quad ,\qquad G^-_{\mu} = \cdot\,
\left(-g_{\mu}\right)\,,\\
\label{jpeq}
J^+_{\mu} &=& j_{\mu}\,\cdot \quad ,\qquad J^-_{\mu} = \cdot\,
j_{\mu}\,.
\end{eqnarray}
Correspondingly, we write $A_I \!=\! \sum_p\!\! :\!\!(A^p_{I+l}J^p_{+l} +
A^p_{I-l}J^p_{-l})\!\!:$ with
\begin{eqnarray}
\label{ap+eq}
A^+_{I+l} &=& \frac{e}{2}|l\rangle\langle 0| \,\cdot \quad ,\qquad
A^-_{I+l} = \cdot\, \frac{e}{2}|l\rangle\langle 0|\,,\\
\label{ap-eq}
A^+_{I-l} &=& -\frac{e}{2}|0\rangle\langle l| \,\cdot \quad ,\qquad
A^-_{I-l} = \cdot\, \left(-\frac{e}{2}|0\rangle\langle l|\right)\,.
\end{eqnarray}
The trace over the bath degrees
of freedom can then be 
performed by application of Wick's theorem.
In this way one obtains a 
series of terms, where vertices $G^p_{\mu}$ and $A^p_{I\pm l}$ of the local
system are connected
by pair contractions $\gamma^{pp'}_{\mu\mu'}(t)={\rm
Tr}_B\left[J^p_{\mu}J^{p'}_{\mu'}\rho_B^{eq}\right]$ of the bath.
Denoting the sum over all irreducible diagrams, which contain the
current superoperators $A^p_{I\pm l}$, by $\Sigma_I$ then leads to
Eq.~(\ref{iteq}), see Fig.~\ref{itfig}.
Similarly, Eq.~(\ref{pteq}) is obtained by introducing the object
$\Sigma$, which is defined as the sum over
all irreducible diagrams involving only the vertices
$G^p_{\mu}$.

(ii) The objects $\Sigma(z)$ and $\Sigma_I(z)$ are calculated by a
renormalization group procedure. Short time scales of
$\gamma^{pp'}_{\mu\mu'}(t)$ are
integrated out first 
by introducing a short-time cutoff $t_c$ into the correlation
function
$\gamma^{pp'}_{\mu\mu'}(t)\rightarrow\gamma^{pp'}_{\mu\mu'}(t,t_c)$.
In each renormalization-group step, the time scales between $t_c$ and
$t_c+dt_c$ are 
integrated out, starting from $t_c=0$ and ending at $t_c=\infty$.
As a consequence, one generates RG equations for $\Sigma(z)$ ($\Sigma_I(z)$), 
$L_0$, $G^p_{\mu}$, and the boundary vertex operators $A^p_{\mu}$
($A^p_{I\mu}$)and $B^p_{\mu}$ 
(defined as the rightmost and leftmost vertex of the kernel
$\Sigma(z)$ ($\Sigma_I(z)$)), for the explicit form of the RG
equations see the Appendix.

Within the scheme of a perturbative RG analysis, the generation of multiple
vertex superoperators is neglected here, as in Ref.~\cite{hs-koenig,mk-hs_sbm}.
Due to the small realistic value of $g$ ($g=0.05$ for GaAs) the
neglecting of double- and
higher-order vertex objects is justified and our approach leads to very
reliable results, see Ref.~\cite{mk-hs_sbm}, where we solved the spin-boson
model for couplings up to $\alpha=g/2\lesssim 0.1..0.2$. 
Furthermore, since also the coupling to the leads is treated
nonperturbatively, the induced level broadening is accounted for
properly. In contrast, in a perturbative analysis this effect has to be
accounted for by an additional cutoff parameter.
Also note that, in our method there is no restriction regarding the
temperature, whereas the NIBA is only valid for
the parameter regime $(\Delta_r^2+\epsilon^2)^{1/2}\lesssim
T$~\cite{weiss-book}. Here,
$\Delta_r=2T_c(2T_c/D)^{\alpha/(1-\alpha)}$ is the renormalized tunnel
amplitude of the spin-boson model.
Finally, using the RTRG we are also able to account for the offdiagonal
electron-phonon interaction, which is important for small
$D/\omega_d$, see Eqs.~(\ref{Tceq}), (\ref{r22eq}) - (\ref{s23eq}).

{\it Results}.
We solve the set of ordinary differential equations,
Eqs.~(\ref{dsigmaeq}) - (\ref{dbeq}), numerically. The stationary tunnel
current then follows from Eqs.~(\ref{isteq}) and (\ref{psteq}). Our choice
of the parameters corresponds to the experiment, where a GaAs
structure was used at the temperature
$T=23mK=1.98\mu eV$ with an external voltage $V=140\mu
eV$~\cite{fujisawa}. For GaAs we have $c_s=5000m/s$ and
$g=0.05$~\cite{bruus}. The distance between the dots is estimated as
$d=200\cdot 10^{-9}m$~\cite{brandes}, which leads to $\omega_d=16.5\mu
eV$.

The result for $T_c=\Gamma_l=\Gamma_r=1\mu eV$, $D_l=D_r=1meV$ and
$D=100\mu eV$ respectively $D=150\mu eV$, which corresponds to the
parameters studied in
Ref.~\cite{brandes}, is shown in Fig.~\ref{ist_brartrgfig}. The external
voltage $V$ gives rise to a finite stationary tunnel current through
the double dot. The elastic
current can be seen around $\epsilon$ with a width depending on the
coupling to the leads $\Gamma_l=\Gamma_r$ and the internal tunnel
amplitude $T_c$. There the phonons do not
participate in the tunnel process. Due to the
coupling to the phonons
there is also an inelastic current, where phonons are emitted
($\epsilon>0$) respectively absorbed ($\epsilon<0$) during the tunnel
process. For increasing $\Gamma_{l(r)}$ the width of the elastic current grows,
while an increased coupling constant $g$ leads to a larger inelastic
current. The effect of
the finite voltage $V$ can be seen in Fig.~\ref{ist_brartrgfig},
where for $\epsilon>V$ the tunnel current drops to zero.
Furthermore, Fig.~\ref{ist_brartrgfig} shows that the offdiagonal
interaction leads to a larger inelastic
current. This
effect is
increased with decreasing $D$, see also Eqs.~(\ref{r22eq}) -
(\ref{s23eq}). We will see below
(Fig.~\ref{ist_exprtrgfig}) that, due to Eq.~(\ref{Tceq}), the
offdiagonal interaction also
suppresses the elastic current. 
Eventually, in Fig.~\ref{ist_brartrgfig} one also recognizes the oscillations
stemming from
the interference of the phonons interacting with the two dots~\cite{brandes}.

Let us now study the current quantitatively in comparison with the experiment.
For this it is necessary to choose realistic parameter values for
$T_c$, $\Gamma_{l(r)}$, $D$ and
$D_{l(r)}$. From changing the bias polarity in the experiments the
ratio $\Gamma_r/\Gamma_l\approx 0.5..1$ was found~\cite{fujisawa}. To
determine the
couplings $T_c$ and $\Gamma_r$, the
experimental data are compared with the result of Stoof and
Nazarov~\cite{stoof}
\begin{equation}
\label{stoofeq}
I_{st} = \frac{T_c^2\Gamma_r}{T_c^2(2+\Gamma_r/\Gamma_l) + \Gamma_r^2/4 + \epsilon^2}\,,
\end{equation}
which is valid for no electron-phonon interaction ($g=0$).
A good agreement of the elastic current is found for $T_c=0.124\mu
eV$ and $\Gamma_l=\Gamma_r=3.5\mu eV$, see Fig.~\ref{ist_expstooffig}.

However, due to the absent electron-phonon interaction the influence
of the finite extension of the electron
densities within the dots is also disregarded in the Stoof-Nazarov result.
In contrast, our method accounts for this extension by the high-energy
cutoff $D$. From Eq.~(\ref{Tceq}) we see that for a finite $D$ the
tunnel amplitude is effectively reduced. Therefore, the Stoof-Nazarov
result underestimates the
value of $T_c$.

From the large inelastic current in the experiment one can conclude,
that in fact a finite value of $D$ was realized. It turns out that
$T_c\approx 0.375$ allows sensible fits.
First, in Fig.~\ref{ist_exprtrgfig} our results for
$T_c=0.375\mu eV$, $\Gamma_l=\Gamma_r=3.5\mu eV$, $D_l=D_r=1meV$ and
$D=70\mu eV$
respectively $D=100\mu eV$ are compared with the experiment. One 
recognizes that with decreasing $D$ the larger overlap of the dots'
wavefunctions leads to a stronger impact of the offdiagonal
electron-phonon interaction. The elastic current is suppressed, whereas
the inelastic current is increased. The
deviations from the experiment show, that there is an $\epsilon$
dependence of the width of the
electron densities, which we have to account for in order to achieve
agreement. In Fig.~\ref{sig_rtrgfig} we show a fit of the width
$\sigma=d\omega_d/D$, which is based on the experimental results for
$I_{st}$. One
recognizes that for larger absolute values of $\epsilon$ the electron
densities are
more sharply peaked. The asymmetry is due to the finite external
voltage $V$. For $\epsilon<0$ the state $|l\rangle$ lies in a deep
potential well, thus this energetic separation of the two quantum
dot levels and the leads causes a very small overlap of the
wavefunctions within the dots. On the other hand for $\epsilon>0$
neither dot level
lies in a deep potential well, however, an increasing energetic
separation $\epsilon$ again leads to more sharply defined electron densities.

In Fig.~\ref{ist_exprtrgfig} one also recognizes that the structure on
the emission side ($\epsilon>0$) of the current spectrum observed in
the experiment does not stem from interference effects of the
phonons. In fact, the oscillations generated by this interference
occur on a much larger energy scale than the structure found in the
experimental curve. In contrast our results show that the bump on the
emission side of the current spectrum is due to another
mechanism: on the one hand we see in Fig.~\ref{ist_exprtrgfig}
that for small $\epsilon>0$ the inelastic
current grows with increasing $\epsilon$. On the other hand however,
the electron densities are simultaneously sharpened, so that the
cutoff $D$ is increased. For larger $\epsilon$ this again reduces the
inelastic current.

In summary, we have applied the RTRG method to the coupled quantum dot
system in a phonon bath in nonequilibrium. By accounting for 
both the coupling to the leads and the coupling to the environmental
phonons nonperturbatively we achieved a
reliable solution for
the stationary tunnel current. For the first time both the elastic and
the inelastic current of the experiment could
quantitatively be reproduced. Our analysis shows the importance of
the finite width $\sigma$ of the electron densities within one dot,
and for the experiment, the dependence of $\sigma$ on the energy
difference $\epsilon$
between the dots was calculated.

{\it Acknowledgments}.
We acknowledge useful discussions with T. Brandes and
K. Sch\"onhammer. We also thank
T. Fujisawa for the experimental data. This work was supported by the
``Deutsche
Forschungsgemeinschaft'' as part of ``SFB 345'' (M.K.) and ``SFB 195''
(H.S.).

\vspace{1cm}
\begin{appendix}
\begin{center}{\bf Appendix}\end{center}

To obtain the explicit form of the RG equations one has to make a
choice of the $t_c$ dependence of the bath contractions
$\gamma^{pp'}_{\mu\mu'}(t,t_c)$.
Since the bosonic bath contractions $\gamma^{pp'}_{b_jb_k}(t)$
($j,k=1,2,3$) correspond to those of the spin-boson model, we choose the
$t_c$ dependence as in Refs.~\cite{mk-hs_sbm,mk-hs_sbmproc}:
\begin{eqnarray}
\gamma^{pp'}_{b_jb_k}(t,t_c) &=& \frac{d}{dt}\left({\tilde R}_{jk}(t)\Theta(t-t_c)\right)\nonumber\\
&& + ip'S_{jk}(t)\Theta(t-t_c)\,.
\label{gammabtceq}
\end{eqnarray}
The functions ${\tilde R}_{jk}(t)$ and $S_{jk}(t)$ are defined by
\begin{equation}
\label{gammarseq}
\gamma^{pp'}_{b_jb_k}(t) = \frac{d}{dt}{\tilde R}_{jk}(t) + ip'S_{jk}(t)\,.
\end{equation}
From Eqs.~(\ref{alphaq1eq}) - (\ref{gammaq1eq}) we obtain
\begin{eqnarray}
{\tilde R}_{11}(t) &=& g \,{\rm Re}\Bigl[\pi T\coth\left(\pi
 T(t-i/D)\right)\Bigr.\nonumber\\
&&\Bigl. +
\frac{\omega_d}{2}\ln\left(\frac{\sinh\left(\pi
T(t-1/\omega_d-i/D)\right)}{\sinh\left(\pi
T(t+1/\omega_d-i/D)\right)}\right)\Bigr]\label{r11eq}\,,\\
\label{s11eq}
S_{11}(t) &=& g \,{\rm
Im}\Bigl[\frac{1/\omega_d^2}{\left((t-i/D)^2-1/\omega_d^2\right)\left(t-i/D\right)^2}\Bigr]\,,\\
\label{rs12eq}
{\tilde R}_{12} &=& {\tilde R}_{21} = S_{12} = S_{21} = 0\,,\\
\label{rs13eq}
{\tilde R}_{13} &=& {\tilde R}_{31} = S_{13} = S_{31} = 0\,,\\
\label{r22eq}
{\tilde R}_{22}(t) &=& \frac{g}{2}e^{-D/\omega_d} \,{\rm Re}\Bigl[\pi T\coth\left(\pi T(t-i/D)\right)\Bigr]\,,\\
\label{s22eq}
S_{22}(t) &=& -\frac{g}{2}e^{-D/\omega_d} \,{\rm
Im}\Bigl[\frac{1}{(t-i/D)^2}\Bigr]\,,\\
{\tilde R}_{23}(t) &=& {\tilde R}_{32}(t) = -g \,{\rm
Re}\Bigl[\omega_d e^{-D/2\omega_d} \Bigr.\nonumber\\
&& \Bigl.\times\ln\left(\frac{\sinh\left(\pi
T(t-1/2\omega_d-i/D)\right)}{\sinh\left(\pi
T(t+1/2\omega_d-i/D)\right)}\right)\Bigr]\label{r23eq}\,,\\
\label{s23eq}
S_{23}(t) &=& S_{32}(t) = -g \,{\rm
Im}\Bigl[\frac{e^{-D/2\omega_d}}{\left((t-i/D)^2-1/4\omega_d^2\right)}\Bigr]\,,\\{\tilde R}_{33}(t) &=& g \,{\rm Re}\Bigl[\pi T\coth\left(\pi
 T(t-i/D)\right)\Bigr.\nonumber\\
&&\Bigl. -
\frac{\omega_d}{2}\ln\left(\frac{\sinh\left(\pi
T(t-1/\omega_d-i/D)\right)}{\sinh\left(\pi
T(t+1/\omega_d-i/D)\right)}\right)\Bigr]\label{r33eq}\,,\\
\label{s33eq}
S_{33}(t) &=& g \,{\rm
Im}\Bigl[\frac{1/\omega_d^2 - 2(t-i/D)^2}{\left((t-i/D)^2-1/\omega_d^2\right)\left(t-i/D\right)^2}\Bigr]\,.
\end{eqnarray}
Here, we assumed that the high-energy cutoff $D$ is much larger than
the temperature $T$. In Eqs.~(\ref{r22eq}) - (\ref{s23eq}) one again
recognizes that the influence of the
offdiagonal interaction strongly depends on the width of the electron
densities $\sigma=|\vec d|\omega_d/D$.
The fermionic contractions can be written as
\begin{eqnarray}
\label{gammaffeq}
\gamma^{pp'}_{\eta f\eta'f'}(t) &=& \delta_{\eta,-\eta'}\delta_{f,f'}
\left(\frac{1}{2}\left(\gamma_{\eta f}(t)+\gamma_{-\eta
f}(-t)\right)\right.\nonumber\\
&&\left. +
\frac{p'}{2}\left(\gamma_{\eta f}(t)-\gamma_{-\eta f}(-t)\right)\right)\,,
\end{eqnarray}
where the indices $\eta$ and $f$ run over $\pm$ and $l,r$. 
We introduce
\begin{eqnarray}
\label{gammaleq}
\Gamma_l(\epsilon) &=& 2\pi \sum_k |V_k|^2\delta(\epsilon-\epsilon_k)\,,\\
\label{gammareq}
\Gamma_r(\epsilon) &=& 2\pi \sum_k |W_k|^2\delta(\epsilon-\epsilon_k)\,,
\end{eqnarray}
so that we obtain
\begin{equation}
\label{gammaetafeq}
\gamma_{\eta f}(t) =
\frac{-iT\Gamma_fe^{-i\eta\mu_ft}}{2\sinh\left(\pi
T(t-i/D_f)\right)}\,.
\end{equation}
Here, we introduced a bandwidth $D_f$ of the reservoir $f$ and assumed,
that $\Gamma_f(\epsilon)\approx{\rm const.}$ holds.
The cutoff-dependence of the fermionic contractions is chosen as
\begin{equation}
\label{gammaftceq} 
\gamma^{pp'}_{\eta f\eta'f'}(t,t_c) = \gamma^{pp'}_{\eta
f\eta'f'}(t)\Theta(t-t_c)\,.
\end{equation}

The final RG equations then read (for a more detailed derivation we
refer to Ref.~\cite{hs})
\begin{eqnarray}
\label{dsigmaeq}
\frac{d\Sigma_{(I)}}{dt_c} &=&
 \sum_{p_1,p_2,j,k}-i\left({\tilde R}_{jk}(t_c)A_{(I)b_j}^{p_1}(t_c)(L_0-z)\right.\nonumber\\
&&\left.\hspace{1cm} +
p_2S_{jk}(t_c)A_{(I)b_j}^{p_1}(t_c)\right)B_{b_k}^{p_2}\nonumber\\
&&\hspace{-0.5cm} - \sum_{p_1,p_2,\eta,f}\gamma^{p_1p_2}_{-\eta f\eta
f}(t_c)\hat\sigma^{p_1p_2}A_{(I)-\eta f}^{p_1}(t_c)B_{\eta f}^{p_2}\,,\\
\label{dleq}
\frac{dL_0}{dt_c} &=& \sum_{p_1,p_2,j,k}\left({\tilde R}_{jk}(t_c)[G_{b_j}^{p_1}(t_c),L_0]\right.\nonumber\\
&&\left.\hspace{1cm} +
p_2S_{jk}(t_c)G_{b_j}^{p_1}(t_c)\right)G_{b_k}^{p_2}\nonumber\\
&&\hspace{-0.5cm} - i\sum_{p_1,p_2,\eta,f}\gamma^{p_1p_2}_{-\eta f\eta
f}(t_c)\hat\sigma^{p_1p_2}G_{-\eta f}^{p_1}(t_c)G_{\eta f}^{p_2}\,,\\
\label{dgeq}
\frac{dG_{\mu}^p}{dt_c} &=& \sum_{p_1,p_2,j,k}
{\tilde R}_{jk}(t_c)\left(G_{b_j}^{p_1}(t_c)G_{\mu}^p-G_{\mu}^pG_{b_j}^{p_1}(t_c)\right)G_{b_k}^{p_2}\nonumber\\
&& -
i\int_0^{t_c}dt\,\left(G_{b_j}^{p_1}(t)G_{\mu}^p-G_{\mu}^pG_{b_j}^{p_1}(t)\right)\nonumber\\
&& \times \left({\tilde R}_{jk}(t_c)[L_0,G_{b_k}^{p_2}(t-t_c)]\right.\nonumber\\
&&\left.\hspace{1cm} +
p_2S_{jk}(t_c)G_{b_k}^{p_2}(t-t_c)\right)\nonumber\\
&& + \sum_{p_1,p_2,\eta,f}\gamma^{p_1p_2}_{-\eta f\eta
f}(t_c)\int_0^{t_c}dt\, \left(G_{\mu}^p\hat\sigma^{p_1p_2}G_{-\eta
f}^{p_1}(t)\right.\nonumber\\
&&\left.\hspace{0.0cm} - \eta^{pp_2}_{\mu}\hat\sigma^{p_1p_2}G_{-\eta
f}^{p_1}(t)G_{\mu}^p\right)G_{\eta f}^{p_2}(t-t_c)\,,\\
\label{daeq}
\frac{dA_{(I)\mu}^p}{dt_c} &=& \sum_{p_1,p_2,j,k}
{\tilde R}_{jk}(t_c)\left(A_{(I)b_j}^{p_1}(t_c)G_{\mu}^p\right.\nonumber\\
&&\left.\hspace{1cm} - A_{(I)\mu}^pG_{b_j}^{p_1}(t_c)\right)G_{b_k}^{p_2}\nonumber\\
&& -
i\int_0^{t_c}dt\,\left(A_{(I)b_j}^{p_1}(t)G_{\mu}^p-A_{(I)\mu}^pG_{b_j}^{p_1}(t)\right)\nonumber\\
&& \times \left({\tilde R}_{jk}(t_c)[L_0,G_{b_k}^{p_2}(t-t_c)]\right.\nonumber\\
&&\left.\hspace{1cm} + p_2S_{jk}(t_c)G_{b_k}^{p_2}(t-t_c)\right)\nonumber\\
&& + \sum_{p_1,p_2,\eta,f}\gamma^{p_1p_2}_{-\eta f\eta
f}(t_c)\nonumber\\
&& \times\int_0^{t_c}dt\, \left(A_{(I)\mu}^p\hat\sigma^{p_1p_2}G_{-\eta
f}^{p_1}(t)\right.\nonumber\\
&&\left.\hspace{0.0cm} - \eta^{pp_2}_{\mu}\hat\sigma^{p_1p_2}A_{(I)-\eta
f}^{p_1}(t)G_{\mu}^p\right)G_{\eta f}^{p_2}(t-t_c)\,,\\
\label{dbeq}
\frac{dB_{\mu}^p}{dt_c} &=& \sum_{p_1,p_2,j,k}
{\tilde R}_{jk}(t_c)G_{b_j}^{p_1}(t_c)G_{\mu}^pB_{b_k}^{p_2} -
i\int_0^{t_c}dt\,G_{b_j}^{p_1}(t)\nonumber\\
&& \times  G_{\mu}^p\left({\tilde R}_{jk}(t_c)(L_0-z)B_{b_k}^{p_2}(t-t_c)\right.\nonumber\\
&&\left.\hspace{1cm} + p_2S_{jk}(t_c)B_{b_k}^{p_2}(t-t_c)\right)\nonumber\\
&& - \sum_{p_1,p_2,\eta,f}\gamma^{p_1p_2}_{-\eta f\eta
f}(t_c)\nonumber\\
&& \times\int_0^{t_c}dt\, \eta^{pp_2}_{\mu}\hat\sigma^{p_1p_2}G_{-\eta
f}^{p_1}(t)G_{\mu}^pB_{\eta f}^{p_2}(t-t_c)\,.
\end{eqnarray}
The interaction picture is defined by $G_{\mu}^p(t)=e^{iL_0t}G_{\mu}^p
e^{-iL_0t}$,
$A_{(I)\mu}^p(t)=e^{izt}A_{(I)\mu}^p e^{-iL_0t}$, and
$B_{\mu}^p(t)=e^{iL_0t}B_{\mu}^p e^{-izt}$. 
The function $\eta^{pp'}_{\mu}$ and the superoperator
$\hat\sigma^{pp'}$ account for additional signs arising from the
commutation of fermionic field operators.
They are given by
\begin{eqnarray}
\label{etaeq}
\eta^{pp'}_{\mu} &=& \left\{\begin{array}{cc} -pp' & \mbox{for $\mu$
fermionic}\\
1 & \mbox{else}
\end{array}\right.\,,\\
\label{sigmaeq}
\left(\hat\sigma^{pp'}\right)_{ss',ss'} &=& \left\{\begin{array}{cc}
pp' & \mbox{for $N_s-N_{s'}={\rm odd}$}\\
1 & \mbox{for $N_s-N_{s'}={\rm even}$}
\end{array}\right.\,.
\end{eqnarray}
We also note, that in the limit of large $D_l$ and $D_r$ the first
part of the fermionic contractions in Eq.~(\ref{gammaffeq}) can be
written as
\begin{equation}
\label{gammadeltaeq}
\frac{1}{2}\left(\gamma_{\eta f}(t)+\gamma_{-\eta
f}(-t)\right) = \frac{\Gamma_f}{2}\delta(t)\,.
\end{equation}
Thus, for $D_l,D_r\rightarrow\infty$, this contribution to the
differential equations
Eqs.~(\ref{dsigmaeq}) - (\ref{dbeq}) can be
incorporated in the initial conditions.
\end{appendix}

\begin{figure}
\centerline{\psfig{figure=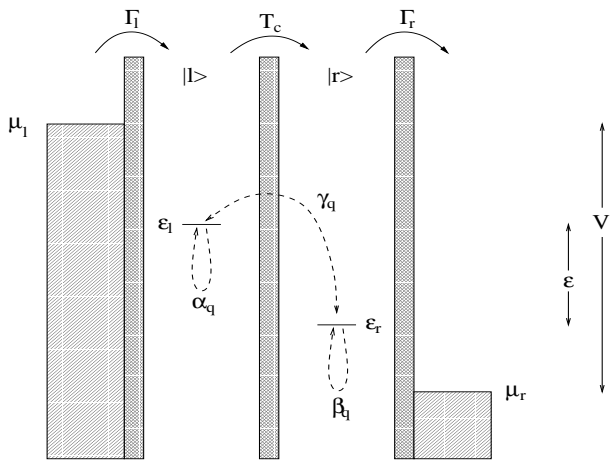,width=8cm,height=6cm,angle=0}}
\vspace{0.5cm}
\caption{The double quantum dot may be charged with one additional
electron in the left or right dot. The corresponding states
$|l\rangle$ and $|r\rangle$ are coupled by the tunnel amplitude
$T_c$. The couplings to the leads are given by $\Gamma_{l(r)}$. The
energy difference between the quantum dots is
$\epsilon=\epsilon_l-\epsilon_r$, and there is an external voltage
$V=\mu_l-\mu_r$. The interaction with the acoustical phonons (dashed
lines), consists of a diagonal part with the coefficients $\alpha_q$
and $\beta_q$, and an
offdiagonal part with the constants $\gamma_q$.}
\label{ddfig}
\end{figure}

\newpage

\begin{figure}
\centerline{\psfig{figure=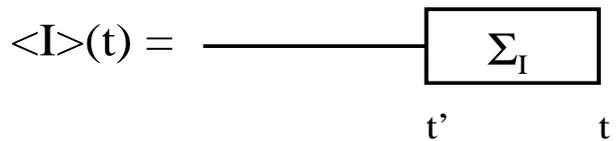,width=8cm,height=1.8cm,angle=0}}
\vspace{0.5cm}
\caption{Diagrammatic expression for $\langle I\rangle (t)$. The two
lines of the Keldysh contour are put together to one line. The
irreducible diagrams in $\Sigma_I$ include the leftmost vertex
superoperator at the time point $t'$ and the current superoperator
$A_I$ at the time point $t$. $\Sigma_I$ acts on $p(t')$, which is
represented by the horizontal line.}
\label{itfig}
\end{figure}

\begin{figure}
\centerline{\psfig{figure=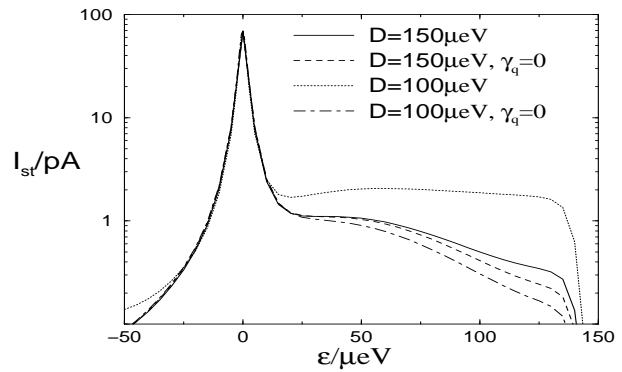,width=8cm,height=5cm,angle=0}}
\caption{Stationary tunnel current as a function of $\epsilon$ for
$T_c=\Gamma_l=\Gamma_r=1\mu eV$ and $D_l=D_r=1meV$. and $D=100\mu eV$
respectively $D=150\mu eV$ . Solid
line: $D=150\mu eV$. Dashed line: $D=150\mu eV$, $\gamma_q=0$. Dotted
line: $D=100\mu eV$. Dot-dashed line: $D=100\mu eV$, $\gamma_q=0$.}
\label{ist_brartrgfig}
\end{figure}

\begin{figure}
\centerline{\psfig{figure=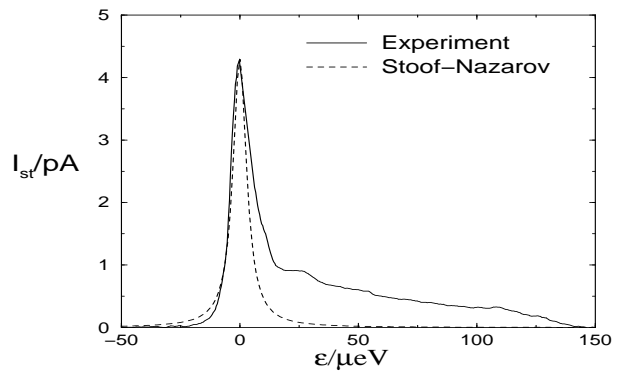,width=8cm,height=5cm,angle=0}}
\caption{Stationary tunnel current as a function of $\epsilon$. Solid
line: Experiment. Dashed line: Stoof-Nazarov result
for the case of no electron-phonon interaction, with the parameters
$T_c=0.124\mu eV$ and $\Gamma_l=\Gamma_r=3.5\mu eV$.}
\label{ist_expstooffig}
\end{figure}

\begin{figure}
\centerline{\psfig{figure=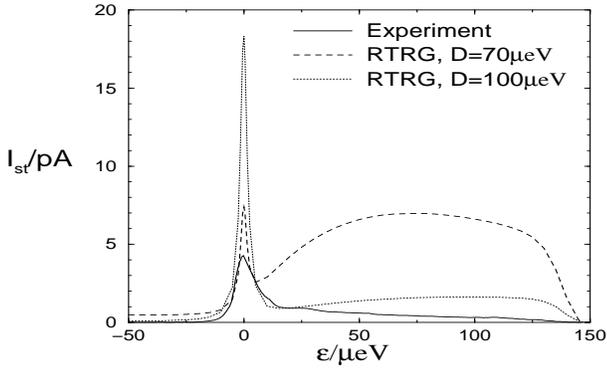,width=7cm,height=5cm,angle=0}}
\caption{Stationary tunnel current as
a function of $\epsilon$. Solid
line: Experiment. Dashed line: RTRG with $T_c=0.375\mu
eV$, $\Gamma_l=\Gamma_r=3.5\mu
eV$, $D_l=D_r=1meV$ and $D=70\mu eV$. Dotted line: RTRG with $T_c=0.375\mu
eV$, $\Gamma_l=\Gamma_r=3.5\mu
eV$, $D_l=D_r=1meV$ and $D=100\mu eV$.}
\label{ist_exprtrgfig}
\end{figure}

\begin{figure}
\centerline{\psfig{figure=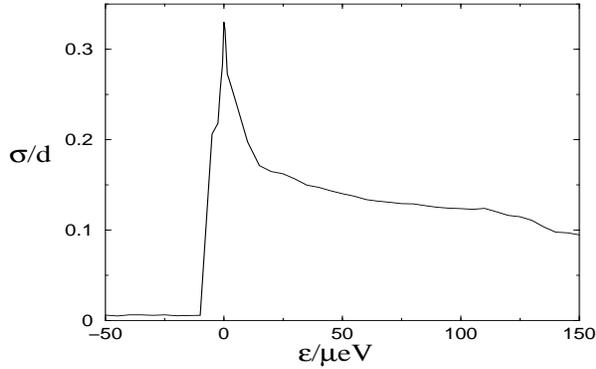,width=8cm,height=5cm,angle=0}}
\caption{The width of the electron density within one dot, $\sigma$, as
a function of $\epsilon$. $T_c=0.375\mu eV$, $\Gamma_l=\Gamma_r=3.5\mu
eV$ and $D_l=D_r=1meV$.}
\label{sig_rtrgfig}
\end{figure}

\end{document}